\documentclass[10pt,a4paper,onecolumn]{article}
\usepackage{marginnote}
\usepackage{graphicx}
\usepackage{xcolor}
\usepackage{authblk,etoolbox}
\usepackage{titlesec}
\usepackage{calc}
\usepackage{tikz}
\usepackage{hyperref}
\hypersetup{colorlinks,breaklinks,
            urlcolor=[rgb]{0.0, 0.5, 1.0},
            linkcolor=[rgb]{0.0, 0.5, 1.0}}
\usepackage{caption}
\usepackage{tcolorbox}
\usepackage{amssymb,amsmath}
\usepackage{ifxetex,ifluatex}
\usepackage{seqsplit}
\usepackage{xstring}

\usepackage{float}
\let\origfigure\figure
\let\endorigfigure\endfigure
\renewenvironment{figure}[1][2] {
    \expandafter\origfigure\expandafter[H]
} {
    \endorigfigure
}

\usepackage{fixltx2e} 

\let\textttOrig=\texttt
\def\texttt#1{\expandafter\textttOrig{\seqsplit{#1}}}
\renewcommand{\seqinsert}{\ifmmode
  \allowbreak
  \else\penalty6000\hspace{0pt plus 0.02em}\fi}

\makeatletter
\let\href@Orig=\href
\def\href@Urllike#1#2{\href@Orig{#1}{\begingroup
    \def\Url@String{#2}\Url@FormatString
    \endgroup}}
\def\href@Notdoi#1#2{\def\tempa{#1}\def\tempb{#2}%
  \ifx\tempa\tempb\relax\href@Urllike{#1}{#2}\else
  \href@Orig{#1}{#2}\fi}
\def\href#1#2{%
  \IfBeginWith{#1}{https://doi.org}%
  {\href@Urllike{#1}{#2}}{\href@Notdoi{#1}{#2}}}
\makeatother

\usepackage[top=3.5cm, bottom=3cm, right=1.5cm, left=1.0cm,
            headheight=2.2cm, reversemp, includemp, marginparwidth=4.5cm]{geometry}

\titleformat{\section}
  {\normalfont\sffamily\Large\bfseries}
  {}{0pt}{}
\titleformat{\subsection}
  {\normalfont\sffamily\large\bfseries}
  {}{0pt}{}
\titleformat{\subsubsection}
  {\normalfont\sffamily\bfseries}
  {}{0pt}{}
\titleformat*{\paragraph}
  {\sffamily\normalsize}

\usepackage{fancyhdr}
\makeatletter
\let\ps@plain\ps@fancy
\fancyheadoffset[L]{4.5cm}
\fancyfootoffset[L]{4.5cm}

\definecolor{linky}{rgb}{0.0, 0.5, 1.0}

\newtcolorbox{repobox}
   {colback=red, colframe=red!75!black,
     boxrule=0.5pt, arc=2pt, left=6pt, right=6pt, top=3pt, bottom=3pt}

\newcommand{\ExternalLink}{%
   \tikz[x=1.2ex, y=1.2ex, baseline=-0.05ex]{%
       \begin{scope}[x=1ex, y=1ex]
           \clip (-0.1,-0.1)
               --++ (-0, 1.2)
               --++ (0.6, 0)
               --++ (0, -0.6)
               --++ (0.6, 0)
               --++ (0, -1);
           \path[draw,
               line width = 0.5,
               rounded corners=0.5]
               (0,0) rectangle (1,1);
       \end{scope}
       \path[draw, line width = 0.5] (0.5, 0.5)
           -- (1, 1);
       \path[draw, line width = 0.5] (0.6, 1)
           -- (1, 1) -- (1, 0.6);
       }
   }

\patchcmd{\@maketitle}{center}{flushleft}{}{}
\patchcmd{\@maketitle}{center}{flushleft}{}{}
\patchcmd{\@maketitle}{\LARGE}{\LARGE\sffamily}{}{}
\def\maketitle{{%
  
  \AB@maketitle}}
\makeatletter
\renewcommand\AB@affilsepx{ \protect\Affilfont}
\renewcommand\AB@affilnote[1]{{\bfseries #1}\hspace{3pt}}
\renewcommand{\affil}[2][]%
   {\newaffiltrue\let\AB@blk@and\AB@pand
      \if\relax#1\relax\def\AB@note{\AB@thenote}\else\def\AB@note{#1}%
        \setcounter{Maxaffil}{0}\fi
        \begingroup
        \let\href=\href@Orig
        \let\texttt=\textttOrig
        \let\protect\@unexpandable@protect
        \def\thanks{\protect\thanks}\def\footnote{\protect\footnote}%
        \@temptokena=\expandafter{\AB@authors}%
        {\def\\{\protect\\\protect\Affilfont}\xdef\AB@temp{#2}}%
         \xdef\AB@authors{\the\@temptokena\AB@las\AB@au@str
         \protect\\[\affilsep]\protect\Affilfont\AB@temp}%
         \gdef\AB@las{}\gdef\AB@au@str{}%
        {\def\\{, \ignorespaces}\xdef\AB@temp{#2}}%
        \@temptokena=\expandafter{\AB@affillist}%
        \xdef\AB@affillist{\the\@temptokena \AB@affilsep
          \AB@affilnote{\AB@note}\protect\Affilfont\AB@temp}%
      \endgroup
       \let\AB@affilsep\AB@affilsepx
}
\makeatother

\renewcommand\Affilfont{\sffamily\small\mdseries}
\ifnum 0\ifxetex 1\fi\ifluatex 1\fi=0 
  \usepackage[T1]{fontenc}
  \usepackage[utf8]{inputenc}

\else 
  \ifxetex
    \usepackage{mathspec}
  \else
    \usepackage{fontspec}
  \fi
  \defaultfontfeatures{Ligatures=TeX,Scale=MatchLowercase}

\fi
\IfFileExists{upquote.sty}{\usepackage{upquote}}{}
\IfFileExists{microtype.sty}{%
\usepackage{microtype}
\UseMicrotypeSet[protrusion]{basicmath} 
}{}

\usepackage{hyperref}
\hypersetup{unicode=true,
            pdftitle={StarEstate: A Python Package for Galactic Population Synthesis},
            pdfborder={0 0 0},
            breaklinks=true}
\usepackage{natbib}
\usepackage{longtable,booktabs}

\let\addcontentslineOrig=\addcontentsline
\def\addcontentsline#1#2#3{\bgroup
  \let\texttt=\textttOrig\addcontentslineOrig{#1}{#2}{#3}\egroup}
\let\markbothOrig\markboth
\def\markboth#1#2{\bgroup
  \let\texttt=\textttOrig\markbothOrig{#1}{#2}\egroup}
\let\markrightOrig\markright
\def\markright#1{\bgroup
  \let\texttt=\textttOrig\markrightOrig{#1}\egroup}

\usepackage{graphicx,grffile}
\makeatletter
\def\maxwidth{\ifdim\Gin@nat@width>\linewidth\linewidth\else\Gin@nat@width\fi}
\def\maxheight{\ifdim\Gin@nat@height>\textheight\textheight\else\Gin@nat@height\fi}
\makeatother
\setkeys{Gin}{width=\maxwidth,height=\maxheight,keepaspectratio}
\IfFileExists{parskip.sty}{%
\usepackage{parskip}
}{
\setlength{\parindent}{0pt}
\setlength{\parskip}{6pt plus 2pt minus 1pt}
}
\providecommand{\tightlist}{%
  \setlength{\itemsep}{0pt}\setlength{\parskip}{0pt}}
\ifx\paragraph\undefined\else
\let\oldparagraph\paragraph
\renewcommand{\paragraph}[1]{\oldparagraph{#1}\mbox{}}
\fi
\ifx\subparagraph\undefined\else
\let\oldsubparagraph\subparagraph
\renewcommand{\subparagraph}[1]{\oldsubparagraph{#1}\mbox{}}
\fi

\title{StarEstate: A Python Package for Galactic Population Synthesis}

        \author[1, 2, 3]{Amedeo Romagnolo}

      \affil[1]{Universität Heidelberg, Zentrum für Astronomie (ZAH),
Institut für Theoretische Astrophysik, Albert Ueberle Str. 2, 69120,
Heidelberg, Germany}
      \affil[2]{Dipartimento di Fisica e Astronomia Galileo Galilei,
Università di Padova, Vicolo dell'Osservatorio 3, I--35122 Padova,
Italy}
      \affil[3]{Department of Astronomy and Astrophysics, University of
California, San Diego, La Jolla, CA 92093, USA}
  \date{\vspace{-5ex}}

\begin{document}
\maketitle

\marginpar{
  \sffamily\small

  {\bfseries DOI:} \href{https://doi.org/}{\color{linky}{}}

  \vspace{2mm}

  {\bfseries Software}
  \begin{itemize}
    \setlength\itemsep{0em}
    \item \href{}{\color{linky}{Review}} \ExternalLink
    \item \href{https://github.com/AmedeoRom/StarEstate}{\color{linky}{Repository}} \ExternalLink
    \item \href{}{\color{linky}{Archive}} \ExternalLink
  \end{itemize}

  \vspace{2mm}

  {\bfseries Submitted:} \\
  {\bfseries Published:}

  \vspace{2mm}
  {\bfseries License}\\
  Authors of papers retain copyright and release the work under a Creative Commons Attribution 4.0 International License (\href{https://creativecommons.org/licenses/by/4.0/}{\color{linky}{CC BY 4.0}}).
}

\hypertarget{summary}{%
\section{Summary}\label{summary}}

\href{https://github.com/AmedeoRom/StarEstate}{\textbf{StarEstate}} is a
Python package designed to generate synthetic populations of stars for
the Milky Way or elliptical galaxies. Creating mock catalogs is
essential for interpreting observational data and testing stellar
physics theories. StarEstate produces realistic stellar systems by
assigning positions, ages, and chemical compositions based on
statistical distributions and stellar evolution models.

A key feature is the ability to model detailed sub-structures, including
the distinction between thin and thick disks, and the probabilistic
assignment of stars to specific spiral arms based on dynamical
temperature and age.

\hypertarget{statement-of-need}{%
\section{Statement of need}\label{statement-of-need}}

Population synthesis allows the comparison of theoretical formation
models with observational data. However, a significant bottleneck in
existing workflows is the computational cost of generating large stellar
populations. Numerical integration for every star can result in runtime
hours.

StarEstate addresses this by implementing a pre-calculated sampler
workflow. By storing numerical solutions for integrals in dedicated
files, the software reduces the computational time for generating
stellar populations in spiral and elliptical galaxies by orders of
magnitude.

This efficiency allows for rapid iterations, making StarEstate
particularly useful to:

\begin{enumerate}
\def\labelenumi{\arabic{enumi}.}
\tightlist
\item
  Generate massive statistical samples for Galaxy visualization.
\item
  Map synthetic stellar populations from detailed MESA evolutionary
  simulations
  \citep{Paxton_2011, Paxton_2013, Paxton_2015, Paxton_2018, Paxton_2019, Jermyn_2023}
  or rapid SSE/BSE codes based on \citet{Hurley_2000} equations.
\item
  Analyze stellar types (e.g., OB-type, Wolf-Rayet, Red Supergiant)
  across different galactic environments and input stellar physics.
\end{enumerate}

\hypertarget{methodology}{%
\section{Methodology}\label{methodology}}

\hypertarget{stellar-integration-and-quantization}{%
\subsection{Stellar Integration and
Quantization}\label{stellar-integration-and-quantization}}

A quantization module bins the continuous masses and metallicities drawn
from samplers to the closest matching tracks in a user's library. It
automatically assigns stellar types following the Morgan-Keenan
classification (in combination with \citet{Hurley_2000} evolutionary
types for SSE/BSE) and identifies evolutionary phases based on
configurable physical conditions.

\hypertarget{galactic-morphology}{%
\subsection{Galactic Morphology}\label{galactic-morphology}}

StarEstate supports two galaxy shapes: \textbf{Elliptical Galaxies}
(uniform distribution of radial angles) and \textbf{Spiral Galaxies}
(stars assigned membership probability to specific Milky Way arms).

A novel feature is the use of ``Dynamical Temperature'' (velocity
dispersion) to determine how tightly populations trace spiral arms based
on age \citep{Mackereth_2019}:

\begin{itemize}
\tightlist
\item
  \textbf{Young Tracers (\(< 100\) Myr):} Dynamically ``cold''
  populations (e.g., OB stars) tightly confined to spiral arms.
\item
  \textbf{Intermediate Tracers (\(100\) Myr -- \(1\) Gyr):} Dynamically
  ``warmer'' populations (e.g., Cepheids) showing a more dispersed
  concentration.
\item
  \textbf{Old Tracers (\(> 1\) Gyr):} Dynamically ``hot'' populations
  with large velocity dispersions, tracing a smoother, lower-contrast
  pattern.
\end{itemize}

The spiral arm parameters (below) combine data from various sources to
map the Milky Way's major arms and are all customizable.

\footnotesize
\setlength{\tabcolsep}{3pt}

\begin{longtable}[]{@{}lccccc@{}}
\toprule
\begin{minipage}[b]{(\columnwidth - 5\tabcolsep) * \real{0.14}}\raggedright
Arm Name\strut
\end{minipage} &
\begin{minipage}[b]{(\columnwidth - 5\tabcolsep) * \real{0.17}}\centering
Pitch Angle (\(p\)) {[}\(^\circ\){]}\strut
\end{minipage} &
\begin{minipage}[b]{(\columnwidth - 5\tabcolsep) * \real{0.17}}\centering
Ref. Radius (\(R_{\text{ref}}\)) {[}kpc{]}\strut
\end{minipage} &
\begin{minipage}[b]{(\columnwidth - 5\tabcolsep) * \real{0.17}}\centering
Ref. Angle (\(\beta_{\text{kink}}\)) {[}\(^\circ\){]}\strut
\end{minipage} &
\begin{minipage}[b]{(\columnwidth - 5\tabcolsep) * \real{0.17}}\centering
Radial Range {[}kpc{]}\strut
\end{minipage} &
\begin{minipage}[b]{(\columnwidth - 5\tabcolsep) * \real{0.17}}\centering
Prob.\strut
\end{minipage}\tabularnewline
\midrule
\endhead
\begin{minipage}[t]{(\columnwidth - 5\tabcolsep) * \real{0.14}}\raggedright
\textbf{Scutum-Centaurus}\strut
\end{minipage} &
\begin{minipage}[t]{(\columnwidth - 5\tabcolsep) * \real{0.17}}\centering
\(2.0^{1}\)\strut
\end{minipage} &
\begin{minipage}[t]{(\columnwidth - 5\tabcolsep) * \real{0.17}}\centering
\(3.14^{1, 2}\)\strut
\end{minipage} &
\begin{minipage}[t]{(\columnwidth - 5\tabcolsep) * \real{0.17}}\centering
\(25^{1}\)\strut
\end{minipage} &
\begin{minipage}[t]{(\columnwidth - 5\tabcolsep) * \real{0.17}}\centering
\(3.0 - 16.0^{1, 3}\)\strut
\end{minipage} &
\begin{minipage}[t]{(\columnwidth - 5\tabcolsep) * \real{0.17}}\centering
0.30\strut
\end{minipage}\tabularnewline
\begin{minipage}[t]{(\columnwidth - 5\tabcolsep) * \real{0.14}}\raggedright
\textbf{Sagittarius-Carina}\strut
\end{minipage} &
\begin{minipage}[t]{(\columnwidth - 5\tabcolsep) * \real{0.17}}\centering
\(13.1^{4}\)\strut
\end{minipage} &
\begin{minipage}[t]{(\columnwidth - 5\tabcolsep) * \real{0.17}}\centering
\(4.93^{4}\)\strut
\end{minipage} &
\begin{minipage}[t]{(\columnwidth - 5\tabcolsep) * \real{0.17}}\centering
\(-45^{4}\)\strut
\end{minipage} &
\begin{minipage}[t]{(\columnwidth - 5\tabcolsep) * \real{0.17}}\centering
\(4.0 - 16.0^{1, 3}\)\strut
\end{minipage} &
\begin{minipage}[t]{(\columnwidth - 5\tabcolsep) * \real{0.17}}\centering
0.25\strut
\end{minipage}\tabularnewline
\begin{minipage}[t]{(\columnwidth - 5\tabcolsep) * \real{0.14}}\raggedright
\textbf{Perseus}\strut
\end{minipage} &
\begin{minipage}[t]{(\columnwidth - 5\tabcolsep) * \real{0.17}}\centering
\(9.5^{5}\)\strut
\end{minipage} &
\begin{minipage}[t]{(\columnwidth - 5\tabcolsep) * \real{0.17}}\centering
\(9.94^{6}\)\strut
\end{minipage} &
\begin{minipage}[t]{(\columnwidth - 5\tabcolsep) * \real{0.17}}\centering
\(150^{1}\)\strut
\end{minipage} &
\begin{minipage}[t]{(\columnwidth - 5\tabcolsep) * \real{0.17}}\centering
\(6.0 - 18.0^{1}\)\strut
\end{minipage} &
\begin{minipage}[t]{(\columnwidth - 5\tabcolsep) * \real{0.17}}\centering
0.20\strut
\end{minipage}\tabularnewline
\begin{minipage}[t]{(\columnwidth - 5\tabcolsep) * \real{0.14}}\raggedright
\textbf{Sagittarius-Carina}\strut
\end{minipage} &
\begin{minipage}[t]{(\columnwidth - 5\tabcolsep) * \real{0.17}}\centering
\(13.0^{4}\)\strut
\end{minipage} &
\begin{minipage}[t]{(\columnwidth - 5\tabcolsep) * \real{0.17}}\centering
\(4.0^{7}\)\strut
\end{minipage} &
\begin{minipage}[t]{(\columnwidth - 5\tabcolsep) * \real{0.17}}\centering
\(-100^{1}\)\strut
\end{minipage} &
\begin{minipage}[t]{(\columnwidth - 5\tabcolsep) * \real{0.17}}\centering
\(3.5 - 20^{1, 6, 8}\)\strut
\end{minipage} &
\begin{minipage}[t]{(\columnwidth - 5\tabcolsep) * \real{0.17}}\centering
0.15\strut
\end{minipage}\tabularnewline
\begin{minipage}[t]{(\columnwidth - 5\tabcolsep) * \real{0.14}}\raggedright
\textbf{Local (Orion Spur)}\strut
\end{minipage} &
\begin{minipage}[t]{(\columnwidth - 5\tabcolsep) * \real{0.17}}\centering
\(10.1^{1}\)\strut
\end{minipage} &
\begin{minipage}[t]{(\columnwidth - 5\tabcolsep) * \real{0.17}}\centering
\(8.15^{9}\)\strut
\end{minipage} &
\begin{minipage}[t]{(\columnwidth - 5\tabcolsep) * \real{0.17}}\centering
\(0^{10}\)\strut
\end{minipage} &
\begin{minipage}[t]{(\columnwidth - 5\tabcolsep) * \real{0.17}}\centering
\(6.0 - 9.0^{11}\)\strut
\end{minipage} &
\begin{minipage}[t]{(\columnwidth - 5\tabcolsep) * \real{0.17}}\centering
0.10\strut
\end{minipage}\tabularnewline
\bottomrule
\end{longtable}

\normalsize
\setlength{\tabcolsep}{6pt}

\textbf{Table Notes:} 1. \citet{Reid_2019} 2. Consistent with models
where arms begin near center. 3. Starts near central bar end, extends
into disk. 4. \citet{Vallee_2017} 5. Average around curvature kink in
\citet{Reid_2019}. 6. \citet{Bobylev_2013} 7. ``Kink'' radius of 4.46
kpc \citep{Reid_2019}. 8. Combined inner Norma and distant Outer arm. 9.
Sun Distance. 10. Galactocentric coordinate system. 11. \citet{Xu_2016}

\begin{figure}
\centering
\includegraphics{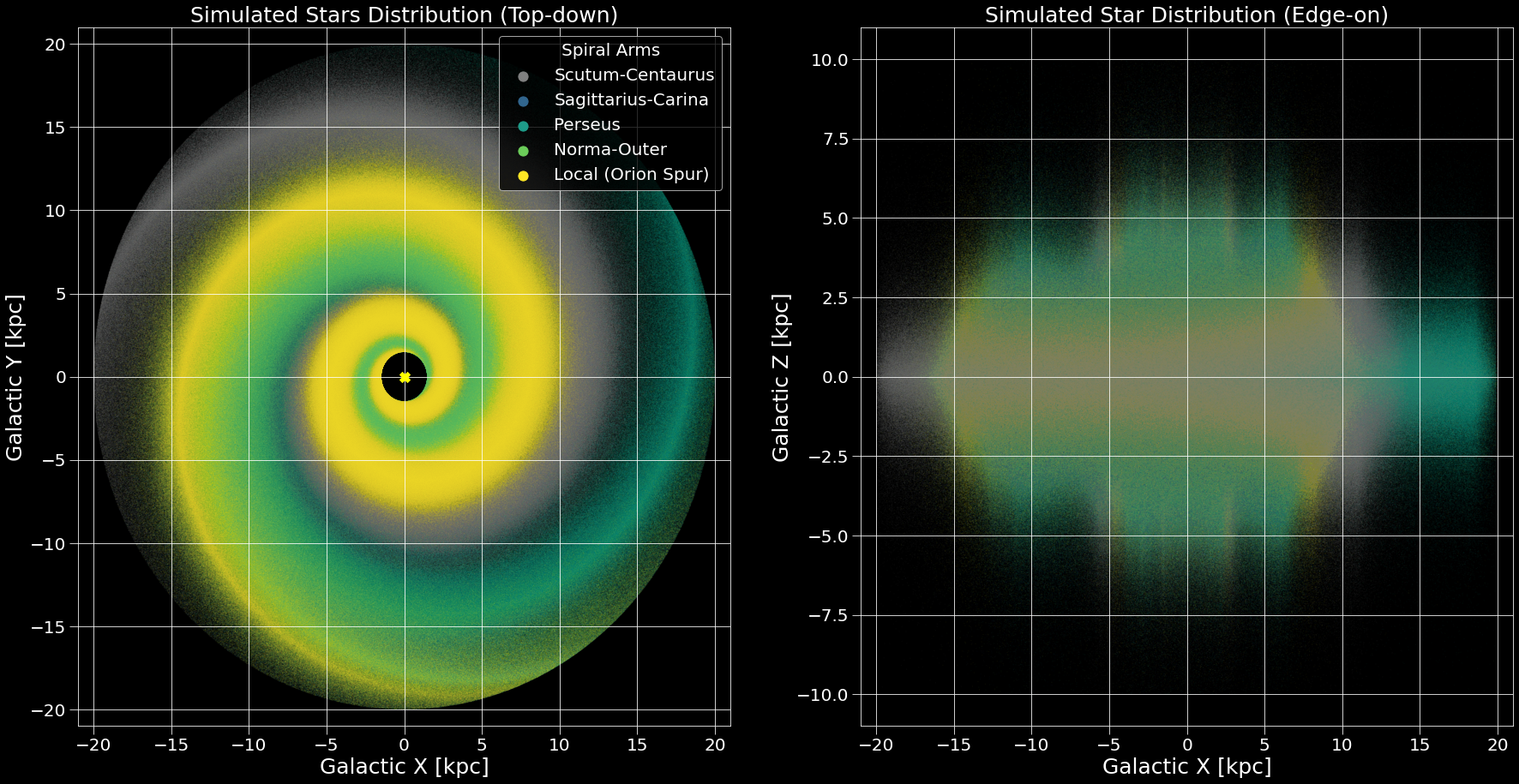}
\caption{Visual distribution of a generated spiral galaxy disk
population (\(N=10^8\) stars), face-on view (left) and edge-on (right).}
\end{figure}

\hypertarget{acknowledgements}{%
\section{Acknowledgements}\label{acknowledgements}}

I acknowledge contributions from Tom Wagg and Floor Broekgaarden. I
acknowledge financial support from the European Research Council for the
ERC Consolidator grant DEMOBLACK, under contract no. 770017 and from the
German Excellence Strategy via the Heidelberg Cluster of Excellence (EXC
2181 - 390900948) STRUCTURES.

\renewcommand\refname{References}
\bibliography{paper.bib}

@article{Hurley_2000,
    author = {Hurley, Jarrod R. and Pols, Onno R. and Tout, Christopher A.},
    title = "{Comprehensive analytic formulae for stellar evolution as a function of mass and metallicity}",
    journal = {Monthly Notices of the Royal Astronomical Society},
    volume = {315},
    number = {3},
    pages = {543-569},
    year = {2000},
    month = {07},
    issn = {0035-8711},
    doi = {10.1046/j.1365-8711.2000.03426.x},
    url = {https://doi.org/10.1046/j.1365-8711.2000.03426.x},
    eprint = {https://academic.oup.com/mnras/article-pdf/315/3/543/2936345/315-3-543.pdf},
}

@article{Reid_2019,
  doi = {10.3847/1538-4357/ab4a11},
  year = {2019},
  publisher = {The American Astronomical Society},
  volume = {885},
  number = {2},
  pages = {131},
  author = {Reid, M. J. and Menten, K. M. and Brunthaler, A. and others},
  title = {Trigonometric Parallaxes of High-mass Star-forming Regions},
  journal = {The Astrophysical Journal}
}

@ARTICLE{Paxton_2011,
  author = {{Paxton}, B. and {Bildsten}, L. and {Dotter}, A. and others},
  title = {{Modules for Experiments in Stellar Astrophysics (MESA)}},
  journal = {ApJs},
  year = {2011},
  volume = {192},
  pages = {3},
  doi = {10.1088/0067-0049/192/1/3}
}

@article{Paxton_2013,
  doi = {10.1088/0067-0049/208/1/4},
  year = 2013,
  volume = {208},
  number = {1},
  pages = {4},
  journal = {ApJs},
  author = {{Paxton}, B. and {Marchant}, P. and {Schwab}, J. and others},
  title = {{MODULES} {FOR} {EXPERIMENTS} {IN} {STELLAR} {ASTROPHYSICS} ({MESA})}
}

@ARTICLE{Paxton_2015,
  author = {{Paxton}, B. and {Marchant}, P. and {Schwab}, J. and others},
  title = {{Modules for Experiments in Stellar Astrophysics (MESA): Binaries, Pulsations, and Explosions}},
  journal = {ApJs},
  year = {2015},
  volume = {220},
  pages = {15},
  doi = {10.1088/0067-0049/220/1/15}
}

@ARTICLE{Paxton_2018,
  author = {{Paxton}, B. and {Schwab}, J. and {Bauer}, E.~B. and others},
  title = {{Modules for Experiments in Stellar Astrophysics (MESA): Convective Boundaries, Element Diffusion, and Massive Star Explosions}},
  journal = {ApJs},
  year = {2018},
  volume = {234},
  pages = {34},
  doi = {10.3847/1538-4365/aaa5a8}
}

@ARTICLE{Paxton_2019,
  author = {{Paxton}, B. and {Smolec}, R. and {Schwab}, Josiah and others},
  title = "{Modules for Experiments in Stellar Astrophysics (MESA): Pulsating Variable Stars, Rotation, Convective Boundaries, and Energy Conservation}",
  journal = {ApJs},
  year = "2019",
  volume = {243},
  number = {1},
  pages = {10},
  doi = {10.3847/1538-4365/ab2241}
}

@article{Jermyn_2023,
  doi = {10.3847/1538-4365/acae8d},
  year = {2023},
  publisher = {The American Astronomical Society},
  volume = {265},
  number = {1},
  pages = {15},
  author = {Adam S. Jermyn and Evan B. Bauer and Josiah Schwab and others},
  title = {Modules for Experiments in Stellar Astrophysics (MESA)},
  journal = {The Astrophysical Journal Supplement Series}
}

@ARTICLE{Mackereth_2019,
  author = {{Mackereth}, J. Ted and {Bovy}, Jo and {Leung}, Henry W. and others},
  title = "{Dynamical heating across the Milky Way disc using APOGEE and Gaia}",
  journal = {MNRAS},
  year = 2019,
  volume = {489},
  number = {1},
  pages = {176-195},
  doi = {10.1093/mnras/stz1521}
}

@article{Bobylev_2013,
  author = {Bobylev, V. V. and Bajkova, A. T.},
  title = {The Milky Way spiral structure parameters from data on masers and selected open clusters},
  journal = {MNRAS},
  volume = {437},
  number = {2},
  pages = {1549-1553},
  year = {2013},
  doi = {10.1093/mnras/stt1987}
}

@article{Vallee_2017,
  title={A guided map to the spiral arms in the galactic disk of the Milky Way},
  volume={13},
  number={3-4},
  journal={Astronomical Review},
  publisher={Informa UK Limited},
  author={Vallée, Jacques P.},
  year={2017},
  pages={113--146},
  DOI={10.1080/21672857.2017.1379459}
}

@article{Xu_2016,
  author = {Ye Xu and Mark Reid and Thomas Dame and others},
  title = {The local spiral structure of the Milky Way},
  journal = {Science Advances},
  volume = {2},
  number = {9},
  pages = {e1600878},
  year = {2016},
  doi = {10.1126/sciadv.1600878}
}

\end{document}